\documentclass[10pt,preprint]{revtex4}
\usepackage{color}
\usepackage{graphicx,subfigure}
\usepackage{amssymb}
\usepackage{amsmath}
\usepackage{booktabs}
\usepackage[english]{babel}
\usepackage{lipsum}
\usepackage{setspace}
\usepackage{array,multirow}
\usepackage{tabularx}

\usepackage{graphicx}
\usepackage{dcolumn}
\usepackage{bm}
\usepackage{float}

\usepackage{forest}

\begin{document}


\title{Phase transitions in a decentralized graph-based approach to human language}

\author{Javier Vera}
 \email{javier.vera@pucv.cl}
\affiliation{ Pontificia Universidad Cat\'olica de Valpara\'iso\\
 Valpara\'iso, Chile}%

\author{Felipe Urbina}
 \email{furbinaparada@gmail.com}
\affiliation{
  Centro de Investigaci\'on DAiTA Lab\\ Facultad de Estudios Interdisciplinarios, Universidad Mayor
  \\
 Santiago, Chile
}%

\author{Wenceslao Palma}
\email{wenceslao.palma@pucv.cl}
\affiliation{%
 Escuela de Ingenier\'ia Inform\'atica\\ Pontificia Universidad Cat\'olica de Valpara\'iso\\
 Valpara\'iso, Chile 
}%

\vspace{5cm}
\begin{abstract}
Zipf's law establishes a scaling behavior for word-frequencies in large text corpora. The appearance of Zipfian properties in human language has been previously explained as an optimization problem for the interests of speakers and hearers. On the other hand, human-like vocabularies can be viewed as bipartite graphs. The aim here is double: within a bipartite-graph approach to human vocabularies, to propose a decentralized language game model for the formation of Zipfian properties. To do this, we define a language game, in which a population of artificial agents is involved in idealized linguistic interactions. Numerical simulations show the appearance of a phase transition from an initially disordered state to three possible phases for language formation. Our results suggest that Zipfian properties in language seem to arise partly from decentralized linguistic interactions between agents endowed with bipartite word-meaning mappings. 
\end{abstract}

\maketitle


\section{\label{sec:intro}Introduction}

This letter arises from two intriguing questions about human language. The first question is: To what extent language, and also language evolution, can be viewed as a graph-theoretical problem? Language is an amazing example of a system of interrelated units at different organization scales. Several recent works have stressed indeed the fact that human languages can be viewed language as a (complex) network of interacting parts \cite{CONG2014598,GAO2014579,doi:10.1002/cplx.20305,Seoane2018}. Within the graph-based approach to human language, one may think word-meaning mappings (that is, \textit{vocabularies}) as bipartite graphs, formed by two sets: words and meanings \cite{doi:10.1002/cplx.20305}.\\

The second question is: What is the nature of the language evolution process that affects the shape of graph-based language representations?  To answer this question, we assume that human communication is constrained (at least) by two forces \cite{doi:10.1002/cplx.20305}: one that pushes towards communicative success and another one that faces the trade-off between speaker and hearer efforts. The first force involves simpler decentralized models of linguistic interactions within populations of artificial agents, endowed with minimal human cognitive features, negotiating pieces of a common language: the so-called \textit{language games} \cite{Loreto_2011,steels2015talking,STEELS2017199,boer2011OxfordHandbook2}. In the simplest language game, the \textit{naming game} \cite{Baronchelli_2006,STEELS2011339}, at discrete time step a pair of players (typically one speaker and one hearer) interacts towards agreement on word-meaning associations.\\

Next, we also consider the communication cost to establish word-meaning mappings. G. Zipf referred to the lexical trade-off between two competing pressures, \textit{ambiguity} and \textit{memory}, as the \textit{least effort principle} \cite{Zipf:36,zipf1949human}: speakers prefer to minimize memory costs; whereas, hearers prefer to minimize disambiguation costs. As remarked by several works, an interesting proposal has stated that human-like vocabularies appear as a phase transition at a critical stage for both competing pressures \cite{Cancho788,FerreriCancho2005,FerreriCancho2005b,CANCHO2006242,1742-5468-2007-06-P06009}. The appearance of a drastic stage of competing pressures can be understood moreover as an explanation of the empirical Zipf's law, which establishes a dichotomy between low-memory words (like the word ``the") and low-ambiguity words (like the word ``cat"). Within a statistical point of view, text corpora evidence strong scaling properties in word-frequencies \cite{Altmann2016,1742-5468-2017-1-014002,DBLP:books/degruyter/KAP2005,1367-2630-16-11-113010,Perc2012,Perc2013,Perc2014,Petersen2012}.\\

The main aim is to address a decentralized approach (based on a previous proposal of two authors of this letter \cite{Urbina_2019}) to the emergence of Zipfian properties in a human-like language, while players communicate with each other using bipartite word-meaning mappings. To structurally characterize changes in the system, our methodology is mainly based on a phase transition description, arising from both classical statistical mechanics tools and graph-mining techniques. We run numerical simulations over simple population topologies. We apply graph-mining techniques, particularly a clustering notion for bipartite graphs \cite{latapy2006basic}.\\ 

\section{The model}

\subsection{\label{sec:graphs} Key concepts on (bipartite) graphs}

A \textit{bipartite graph} is a triple $B=(\top,\bot,E)$, where $\top$ and $\bot$ are two mutually disjoint set of nodes, and $E \subseteq \top \times \bot$ is the set of edges of the graph. Here, $\top$ represents the set of \textit{word nodes}, whereas $\bot$ represents the set of \textit{meaning nodes}. We remark that edges only exist between word nodes and meaning nodes. 
A classical useful tool in graph theory is the matrix representation of graphs. Here, we only consider the \textit{adjacency matrix}. Let us denote by $A = (a)_{wm}$ the adjacency matrix for the (bipartite) graph $B$. From the bipartite sets $\top$ and $\bot$, representing respectively word and meaning nodes, we define the rows of $A$ as word nodes, and the columns as meaning nodes, where $(a)_{wm}=1$ if the word $w$ is joined with the meaning $m$, and 0 otherwise.\\

The \textit{neighbors of order 1} of $u\in \top$ are the nodes at distance 1: $N(u)=\{v\in \bot: uv \in E\}$ (if $u \in \bot$ the definition is analogous). Let us denote by $N(N(u))$ the set of nodes at distance 2 from $u$. The degree $d(u)$ of the node $u$ is simply defined by $d(u)=|N(u)|$. We denote by $d^{\max}_W = \max_{w \in W} d(w)$ the maximum degree for word nodes ($\top$). Analogously, $d^{\max}_M = \max_{m \in M} d(m)$ the maximum degree for meaning nodes ($\bot$).\\

The notion of clustering coefficient (in classical graphs) captures the fact that when there is an edge between two nodes they probably have common neighbors. More generally, such notion captures correlations between neighborhoods. Based on this point of view, \cite{latapy2006basic} proposed a clustering coefficient notion for bipartite graphs:\\

\begin{equation}
    cc(u)=\frac{\sum_{v \in N(N(u))} cc(u,v)}{N(N(u))}
\end{equation}

where $cc(u,v)$ is a notion of clustering defined for pairs of nodes (in the same set $\top$ or $\bot$):

\begin{equation}
    cc(u,v)=\frac{|N(u) \cap N(v)|}{|N(u) \cup N(v)|}
\end{equation}

Interestingly, $cc(u,v)$ captures the overlap between the neighborhoods of $u$ and $v$: if $u$ and $v$ do not share neighbors $cc(u,v)=0$; if they have the same neighborhood $cc(u,v)=1$. 

To give an overall overview of bipartite clustering for the graph $B$, the \textit{average bipartite clustering} reads

\begin{equation*}
    c(B)=\frac{1}{|\top|+|\bot|} \sum_{u \in \top \cup \bot} cc(u)
\end{equation*}

\subsection{\label{sec:basic} Basic elements of the language game}

The language game is played by a finite population of participants $P=$\{1,...,p\}, sharing both a set of words $W=\{1,...,n\}$ and a set of meanings $M=\{1,...,m\}$. Each player $k\in P$ is endowed with a graph-based word-meaning mapping $B^k=(\top^k,\bot^k,E^k)$. In our case, $B^k$ is a bipartite graph with two disjoint sets: $\top^k \subseteq W$ (word nodes) and $\bot^k \subseteq M$ (meaning nodes). Each player $k \in P$ only knows its own graph $B^k$. \\

Two technical terms are introduced. First, we say that a player $k \in P$ \textit{knows} the word $w \in W$ if $w \in \top^k$. Clearly, this definition is equivalent to the existence of the edge $wm \in E^k$, for some $m \in \bot^k$. Second, the \textit{ambiguity} of the word $w$, denoted $a(w)$, is defined as its node degree $d(w)$. 

\subsection{\label{sec:rules} Language game rules}

The dynamics of the language game is based on pairwise speaker-hearer interactions at discrete time steps. At $t \geqslant 0$, a pair of players is selected uniformly at random: one plays the role of \textit{speaker} $s$ and the other plays the role of \textit{hearer} $h$, where $s,h \in P$. Each speaker-hearer communicative interaction is defined by two successive steps. The speaker-centered \textbf{STEP 1} involves the selection of a meaning and a word to transmit them. At \textbf{STEP 2}, the hearer receives the word-meaning association and both speaker and hearer behave according to either \textit{repair} or \textit{alignment} strategies. 

\begin{enumerate}
    \item[] \textbf{STEP 1.} To start the communicative interaction, the speaker $s$ selects the topic of the conversation: one meaning $m^* \in M$. To transmit the meaning $m^*$, the speaker needs to choose some word, denoted $w^*$. There are two possibilities for the selection of $w^*$:
    
    \begin{itemize}
        \item \textbf{if} the edge $wm^* \notin E^s$ for any $w \in \top^s$, the speaker chooses (uniformly at random) the word $w^*$  from the set $W$ and adds the edge $w^*m^*$ to the graph $B^s$;
        
        \item \textbf{otherwise}, if $w^*m^* \in E^s$ for some $w^* \in \top^s$, the speaker \textit{calculates} $w^*$ based on its interests, that is, based on its own conflict between \textit{ambiguity} and \textit{memory}.
    \end{itemize}
    
     To calculate $w^*$ for the second case ($w^*m^* \in E^s$), the speaker behaves according to the ambiguity parameter $\wp \in [0,1]$. Let $random \in [0,1]$ be a random number. Then, two actions are possible:
    
    \begin{itemize}
        \item \textbf{if} $random \geqslant \wp$, the speaker calculates $w^*$ as the \textit{least ambiguous word}
        
        \begin{equation*}
            w^*=\min_{w \in \bot^s} a(w)
        \end{equation*}
        \item \textbf{otherwise}, the speaker calculates $w^*$ as the \textit{most ambiguous word}
        
        \begin{equation*}
            w^*=\max_{w \in \bot^s} a(w)
        \end{equation*}
    \end{itemize}
    
    The speaker transmits the word $w^*$ to the hearer.
    
     \item[] \textbf{STEP 2.} The hearer behaves as in the \textit{naming game}. On the one hand, mutual speaker-hearer agreement (if the hearer knows the word $w^*$) involves \textit{alignment} strategies \cite{STEELS2011339}. On the other hand, a speaker-hearer disagreement (if the hearer does not know the word $w^*$) involves a \textit{repair} strategy in order to increase the chance of future agreements (that is, for $t'>t$). More precisely, 
     
     \begin{itemize}
         \item \textbf{if} the hearer knows the word $w^*$, both speaker and hearer remove all edges formed by $wm^*$, where $w$ respectively belongs to $\top^s \setminus \{w^*\}$ and $\top^h \setminus \{w^*\}$.
         
         \item \textbf{otherwise}, the hearer adds the edge $w^*m^*$ to its graph $B^h$. 
     \end{itemize}
    
\end{enumerate}

\section{\label{sec:methods} Methods}

The population of agents is located on the vertices of a complete graph of size $|P|=100$, typically called the \textit{mean field} approximation. For the description of other simple graph topologies, see the caption of Fig. \ref{cc}. The population shares both a set of $n=|W|=128$ words and a set of $m=|M|=128$ meanings. Starting from an initial condition in which each player $k \in P$ is associated to a bipartite graph $B^k$ where $B^k_{ij} = 1$ or $B^k_{ij} = 0$ with probability 0.5 (put differently, for each possible edge $ij$, $i \in W$ and $j \in M$, exists with probability 0.5), the dynamics performs a speaker-hearer interaction at each discrete time step $t \geqslant 0$. The bipartite word-meaning mappings $B^s$ and $B^h$ are then reevaluated according to communicative success. All results consider averages over 10 initial conditions and $3\times 10^5$ time steps. We denote by $t_f$ the final time step. The ambiguity parameter $\wp$ is varied from 0 to 1 with an increment of 1\%.


\section{Results}

\subsection{Three structural phases in language formation}

Two key quantities have been analyzed for different values of $\wp$: the \textit{average population clustering} $cc$,

 $$   cc=\frac{1}{|P|} \sum_{k \in P} cc(B^k)$$

which captures the average correlation between word neighborhoods; and the (effective) lexicon size at time step $t$, $V(t)$, defined as \cite{Cancho788,Urbina_2019}

 $$   V(t)=\frac{1}{n|P|}\sum_{k \in P} |\top^k|$$

where $V(t)=1$ if $|\top^k|=n$, while $V(t)=0$ if $|\top^k|=0$.\\ 

\begin{figure}[!ht]
\centering
\includegraphics[width=0.4\linewidth]{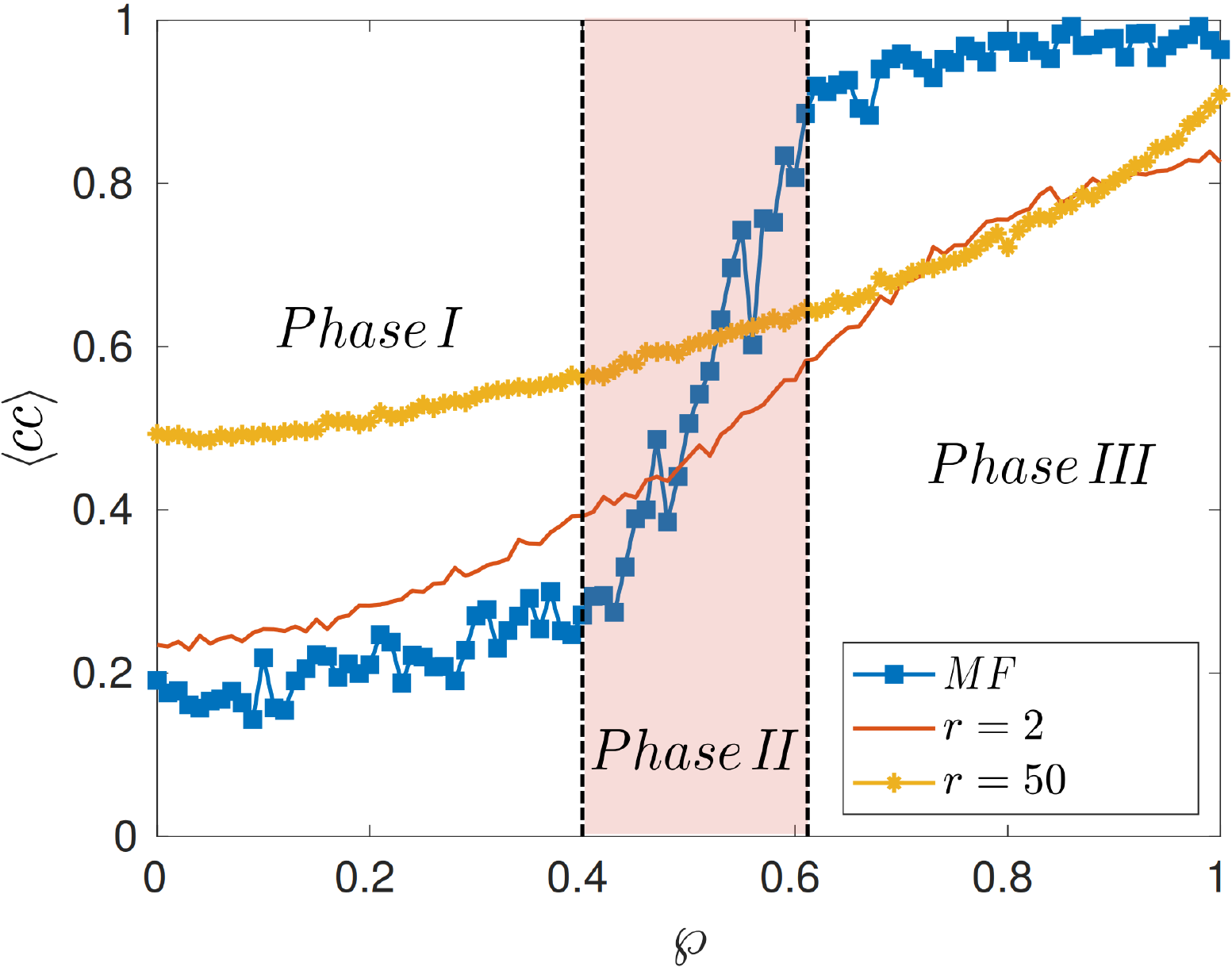}
\caption{\textbf{Average population clustering $\left<cc\right>$ versus ambiguity parameter $\wp$.} It is described the behavior of the average population clustering versus $\wp$. $\langle cc \rangle$ denotes the average over 10 realizations. As shown in the figure, three phases for the evolution of bipartite word-meaning mappings tend to appear: full vocabularies (Phase I), human-like (Phase II) and single-word vocabularies (Phase III). Blue squares indicate the \textit{mean-field} approximation (denoted $MF$). The other two curves indicate a ring topology in which individuals interact at some specific radius $r$. For $r=2$ and $r=50$, the three-phased division of human-like language tends to disappear. This fact suggests that our results are strongly affected by topology features.}
\label{cc}
\end{figure}

\begin{figure}[!ht]
\centering
\includegraphics[width=0.4\linewidth]{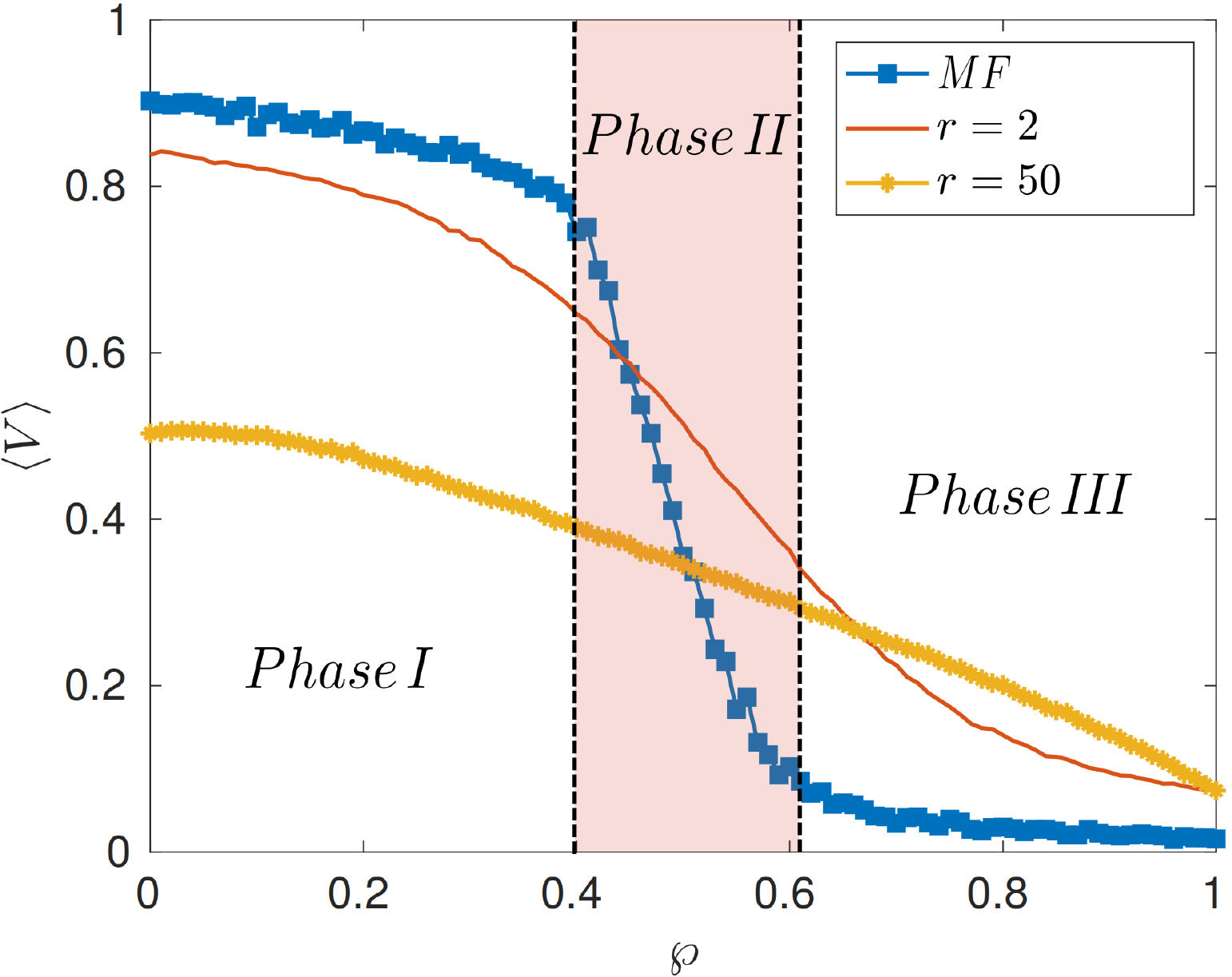}
\caption{\textbf{Effective vocabulary $V(t_f)$ versus ambiguity parameter $\wp$.} It is described the average behavior of $\left<V\right>$ versus $\wp$. Blue squares indicate the \textit{mean-field} approximation (MF). Three phases are exhibited for language formation: full vocabularies (Phase I), human-like (Phase II) and single-word vocabularies (Phase III). }
\label{v}
\end{figure}

Three clear domains can be noticed in the behavior of $\langle cc \rangle$ versus $\wp$, at $t_f$, as shown in Fig. \ref{cc} (blue squares). \textbf{Phase I:} $\langle cc \rangle$ increases smoothly for $\wp < 0.4$, indicating that for this domain there is a small correlation between word neighborhoods. Full vocabularies are attained also for $\wp < 0.4$; \textbf{Phase II:} a drastic transition appears at the critical domain $\wp^* \in (0.4,0.6)$, in which $\langle cc \rangle$ shifts abruptly towards 1. An abrupt change in $V(t_f)$ versus $\wp$ is also found (Fig. \ref{v}) for $\wp^*$; \textbf{Phase III:} single-word languages dominate for $\wp > 0.6$. The maximum value of $\langle cc \rangle$ indicate that word neighborhoods are completely correlated. 


\subsection{\label{sec:vis} Bipartite graphs to visualize the phase transition}

\begin{figure}[!ht]
\centering
\includegraphics[width=0.45\linewidth]{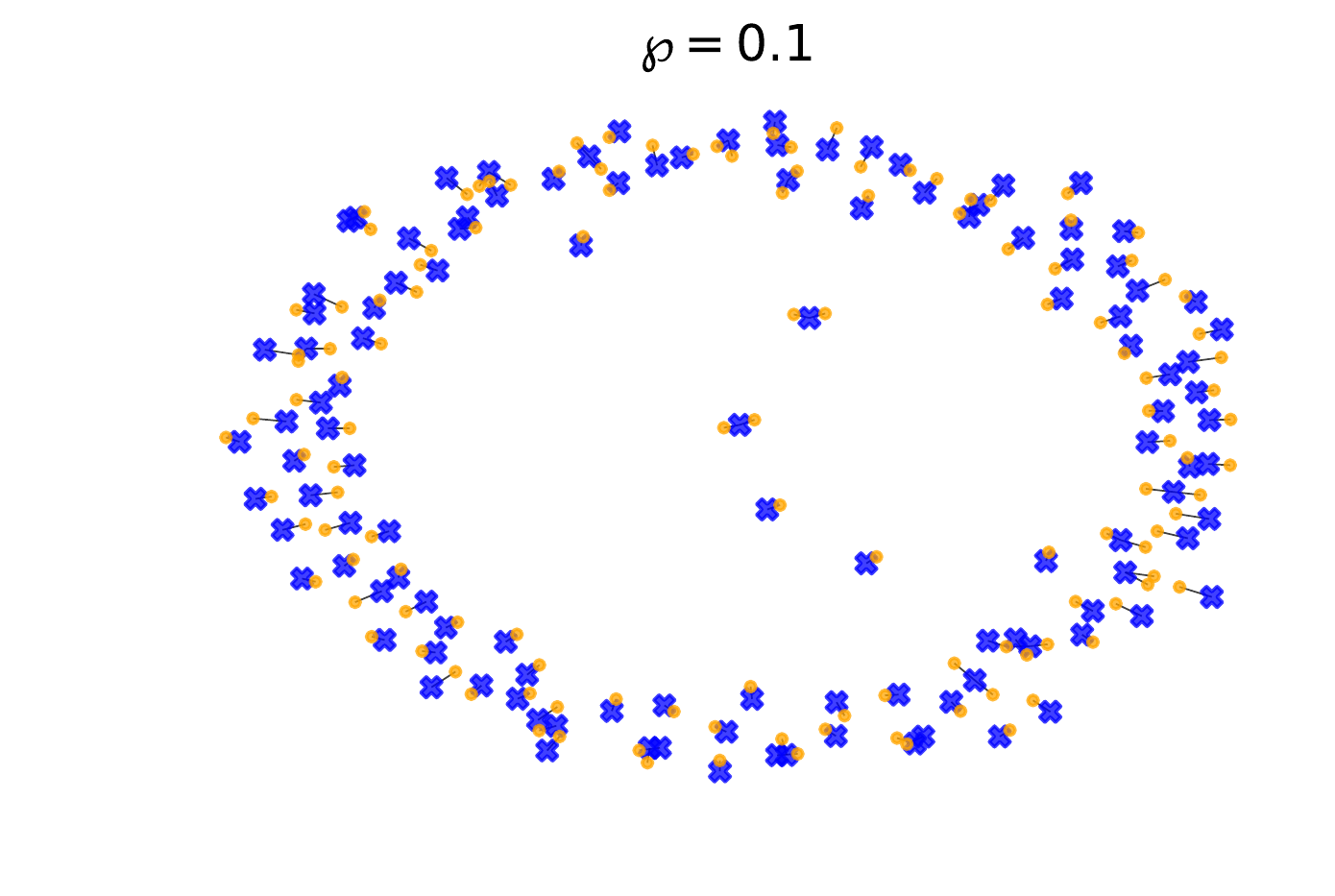}
\includegraphics[width=0.45\linewidth]{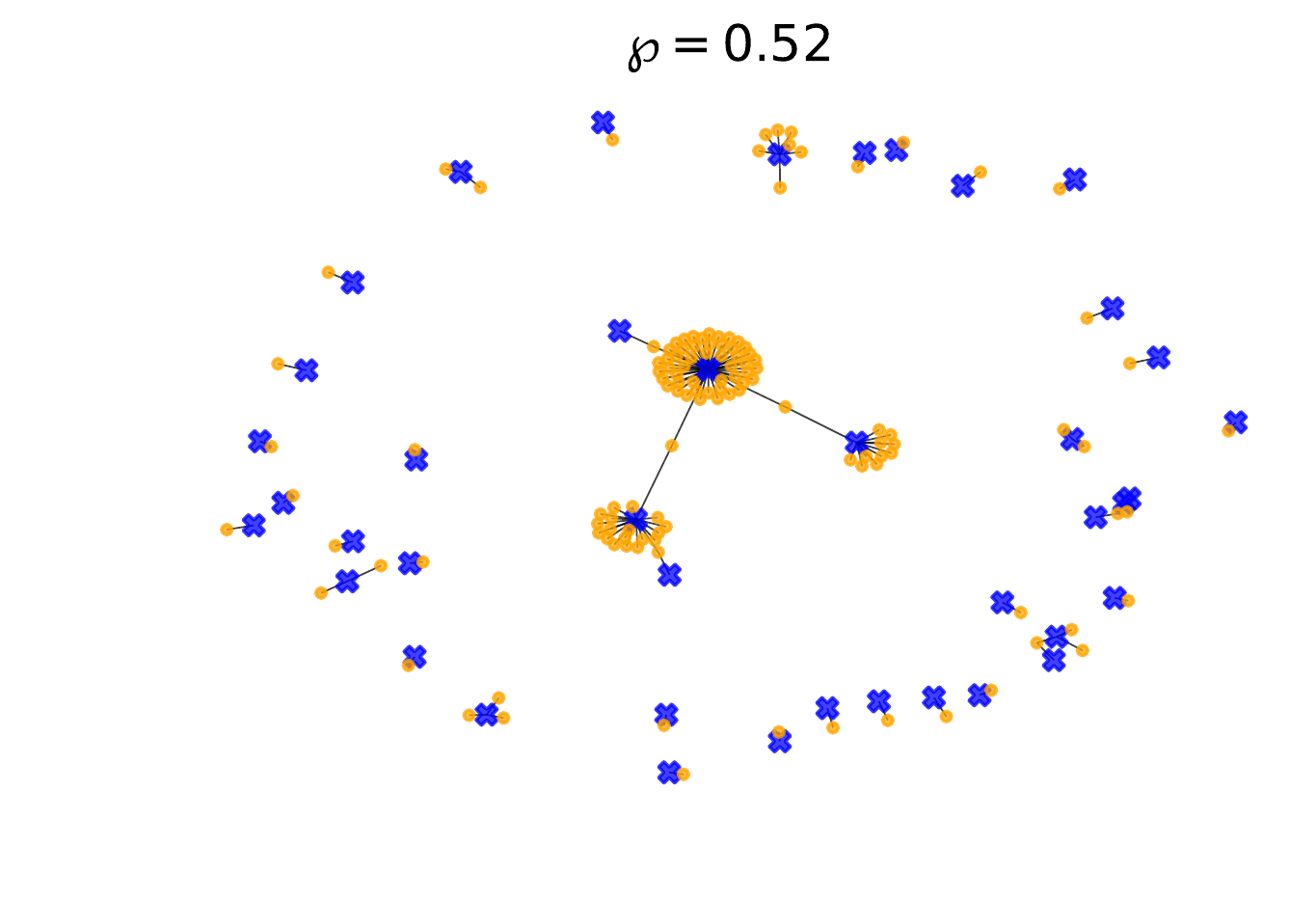}
\includegraphics[width=0.45\linewidth]{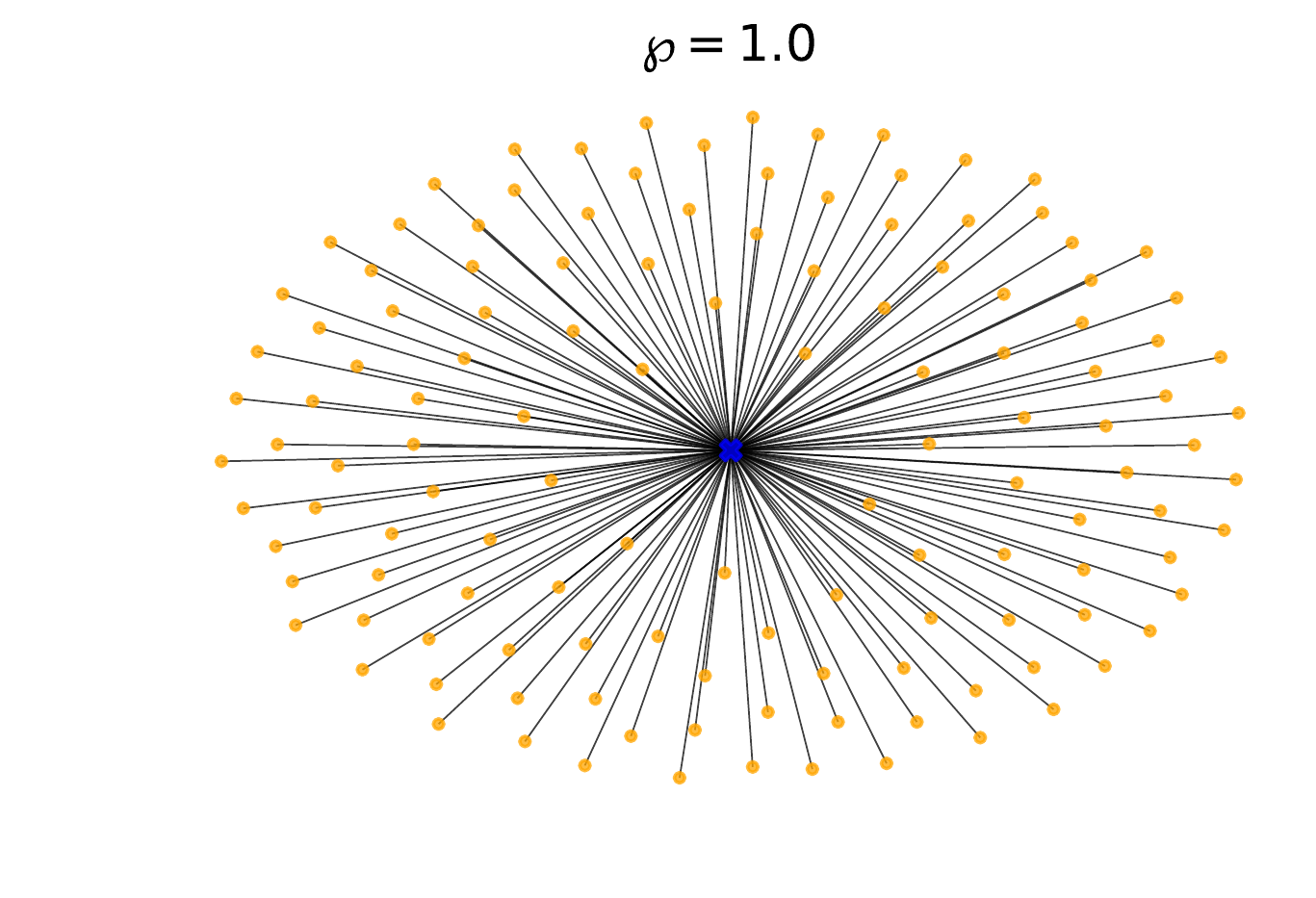}
\caption{\textbf{Visualization of bipartite word-meaning mappings.} Orange circles indicate meanings, whereas blue cruxes indicate words. From top to bottom, we choose three bipartite graphs corresponding respectively to $\wp$ in $\{0.1, 0.52, 1\}$. Node positions are based on a \textit{Python} (https://networkx.github.io/) implementation of the Fruchterman-Reingold algorithm \cite{doi:10.1002/spe.4380211102}.}
\label{graphs}
\end{figure}

We now shift our focus from graph-based measures towards a holistic level in which we illustrate the described phase transition using bipartite graph representations of language formation. We stress the fact that our framework based on a language game with players endowed with bipartite word-meaning mappings is able to visualize the structural changes of the three phases (I, II and III). Fig. \ref{graphs} display, from top to bottom, the bipartite word-meaning mappings for ambiguity parameters $\wp$ in $\{0.1, 0.52,1\}$. As expected, there are radical structural changes between bipartite graphs associated to such ambiguity parameters. Full vocabularies are attained for $\wp=0.1$ (Phase I), located at the hearer-centered phase. Zipfian vocabularies seem to appear for $\wp=0.52$ (Phase II), where speaker and hearer costs have a similar value. Finally, a single-word vocabulary (that is, one word, several meanings) is exhibited for $\wp=1$ (Phase III). 

\newpage
\subsection{\label{sec:trans_energy} Critical values of energy}

The appearance of the three-phased language behavior described here is closely related to previous results of two authors of this letter \cite{Urbina_2019}. Indeed, in the cited paper the energy-like functional $e_{KL}$ (a kullback-leibler-based measure) is minimized around the parameter $\wp\approx 0.5$. Remarkably, here it is showed numerically that around the critical parameter $\wp \approx 0.52$ a drastic transition for both the effective vocabulary and the bipartite average clustering tends to appear (see Fig. \ref{graphs}). 

\begin{figure}[!ht]
\centering
\includegraphics[width=0.4\linewidth]{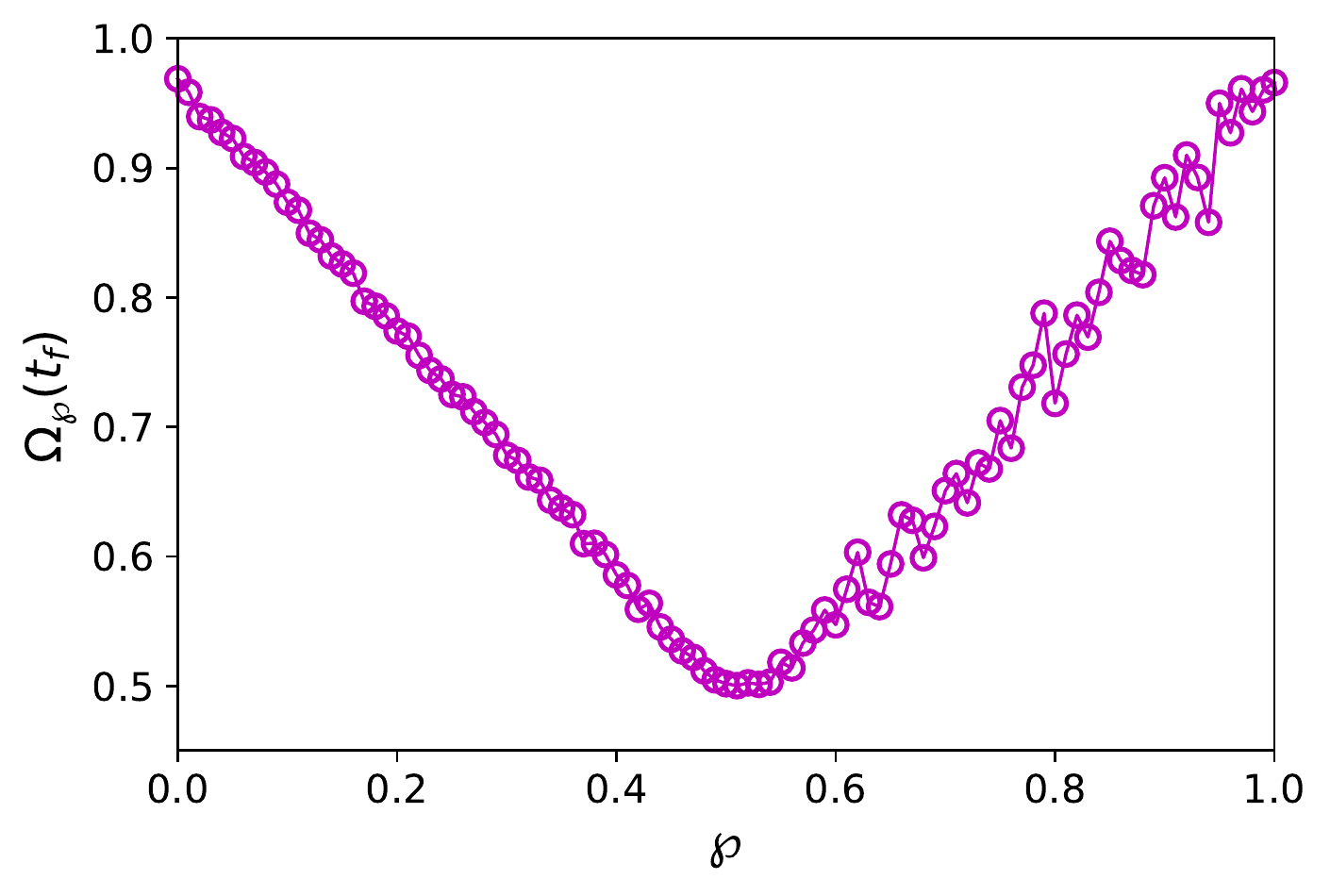}
\caption{\textbf{$\Omega_\wp(t_f)$ versus ambiguity parameter $\wp$.} The figure displays the value of the information-theoretic energy $\Omega_\wp(t_f)$ for one initial condition, after $t_f=3 \times 10^5$ speaker-hearer interactions. $\wp$ is varied with an increment of 1\%.}
\label{omega}
\end{figure}

A first strategy to profound on the problem established between the phase transitions described here and energy-based approaches, is to measure the information-theoretic energy $\Omega_\wp(tf)$ (as defined in \cite{Cancho788}) as a function of the parameter $\wp$. $\Omega_\wp(tf)$ is a combination of the respective efforts of speakers and hearers: $\Omega_\wp(tf)=\wp H(R|S)+(1-\wp)H(S)$. Figure \ref{omega} showed that $\Omega_\wp(t_f)$ is minimized around $\wp\approx 0.5$. This suggests a new way to understand language evolution and formation, by reconciling models focused on self-organization and information-theoretic accounts.  

\section{\label{sec:dis} Discussion}

In this letter, we have described a decentralized model of the emergence of Zipfian features in a human-like language, where agents play language games communicating with bipartite word-meaning mappings. The model evidences a phase transition that corresponds to the formation of a human-like vocabulary satisfying Zipfian word-meaning properties. Our central graph-mining tool has been a notion of clustering for bipartite graphs. This function allowed us to suggest that the drastic transition is, in some sense, a qualitative transition in word's correlations. \\

To further understand the nature of the described transition, we remark a recent proposal \cite{PhysRevLett.122.128301}, reinterpreting an old question about language learning with a novel approach: if language learning by a child involves setting many parameters, to what extent all these need to be innate? According to the \textit{Principles and Parameters} theory \cite{PP}, children are biologically endowed with a general ``grammar" and then the simple exposition to a particular language (for example, \textit{Quechua}) fixes its syntax by equalizing parameters. This debate was illuminated by proposing a statistical mechanics approach in which the distribution of grammar weights (where language is modeled by weighted context-free grammars) evidences a drastic transition. Language learning is, for this proposal, a transition from a random model of grammar parameter-weights to the one in which deep structure (that is, syntax) is encountered.\\

Here, the language learning problem is situated in a decentralized process, with agents negotiating a common word-meaning mapping exhibiting Zipfian scaling properties. Interestingly, our approach can shed light on the debate opened by  \cite{PhysRevLett.122.128301}. Indeed, our model questioned, first, the fact that language learning is traditionally viewed as an individual process, without any consideration of population structure (in general, \textit{language games} question this fact). Secondly, we argue that our view pointed out the minimal necessity of cognitive principles for cultural language formation: the least effort principle. We hypothesize that players only need the most basic cognitive features for language learning (and formation) and the rest is an emergent property from the local speaker-hearer interactions. It is interesting to remark that several works have stressed the fact that language formation can be viewed as a phase transition within an information-theoretic approach \cite{Cancho788,FerreriCancho2005,FerreriCancho2005b,CANCHO2006242,1742-5468-2007-06-P06009}.\\ 

Future work could explore an intriguing hypothesis: Zipfian properties have strong consequences for syntax and symbolic reference. \cite{iCancho2005} has proposed indeed that Zipf's law is a necessary precondition for full syntax, and for going beyond simple word-meaning mappings. They hypothesized moreover that the appearance of syntax have been as abrupt as the transition to Zipf's law. This is a goal for future work: to propose a decentralized model in which agents (constrained by specific cognitive features) develop a Zipfian language that acts as a precondition for the abrupt transition to simple forms of syntax (based, for example, on \cite{Steels2016}). Another related research line arises from models assuming the interplay between maximization of the information transfer and minimization of the entropy of signals (see, for example, \cite{Cancho_2007}). As previously remarked, these models evidence a lack of population structure. Current work asks how a community of individuals playing the language game proposed here can minimize the energy functional $\Omega(\lambda)$ of word-meaning mappings. We may hypothesize that reaching global consensus at the critical phase (that, is for $\wp \approx 0.5$) is closely related to the global minima of $\Omega(\lambda)$ (as a first approach to this problem, see Section \ref{sec:trans_energy}). This idea opens fascinating novel ways to study human language, reconciling models seeing communication as a global minima of information entropic energies and models focused on populations self-organizing themselves towards a shared consensus. 

\begin{acknowledgments}
F.U. thanks CONICYT Chile for financial support under the Grant 3180227. 
\end{acknowledgments}

\bibliographystyle{unsrt}
\bibliography{apssamp}%
\end{document}